\documentclass[journal]{IEEEtran}

\usepackage{relsize}
\usepackage{amsmath}
\usepackage{amssymb}
\usepackage{graphicx}% Include figure files
\usepackage{dcolumn}% Align table columns on decimal point
\usepackage{bm}% bold math

\usepackage{cite}
\usepackage[font=footnotesize]{subfig}
\usepackage{fixltx2e}
\usepackage{stfloats}
\begin{document}
\title{A novel structure designed for high density nonvolatile memory devices}

\author{Jian Cui% <-this % stops a space
\thanks{Email:tsuijian@gmail.com}% <-this % stops a space
}
\maketitle

\begin{abstract}
%\boldmath
The same as in microprocessor fabrication, nonvolatile memory devices are facing the problem in device size scaling down, such as large leakage current density. High-k materials are considered to solve it. However, simply replacing low-k to high-k materials, while keeping the structure as before, is not a good solution. Based on our analysis, we proposed a novel structure, in which charges are injected from top gate electrode. In this structure, high charge injection, large memory window and long retention time can be expected.
\end{abstract}

\IEEEpeerreviewmaketitle

\section{Introduction}
\IEEEPARstart{N}{onvolatile} memory (NVM) has attracted great interests for its enormous applications since its invention. \cite{kahng1967} Large volume, high speed NVM devices are highly demanded in many fields, such as portable memory sticks, smart phones, digital cameras, etc. To achieve large volume and high speed, the size of device approaches to the physical limit. \cite{houdt2011} The capacitance of single devices thus needs to be large enough to maintain channel current large enough to keep a high on-off ratio. If the traditional gate dielectric material, SiO$_2$, is used, it urges the thickness of the dielectric gate to a unprecedented thin level (around or below 1nm). Thus high-k materials are proposed to replace SiO$_2$, which has been utilized in microprocessor fabrication. \cite{mistry2007} In NVM, the common way to use high-k materials is simply change SiO$_2$ to high-k materials. \cite{houdt2005,lu2006} However, we found that in this way, the performance is compromised in traditional floating gate NVM device structure. In this Letter, we proposed a novel NVM structure, in which the charges are injected from top gate electrode. By this way, it is possible to achieve scaling down further without performance compromise.

\section{A contradiction in traditional floating gate NVM}
\label{sec:contra}

\begin{figure}[rt]
\centering
\includegraphics[width=0.35\textwidth]{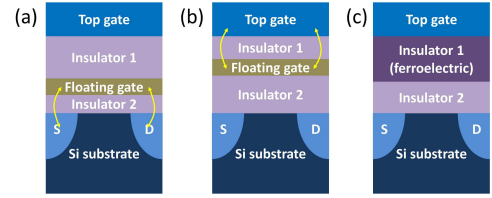}
\caption{(a) Model 1. Floating gate NVM structure with charge injection from substrate or source. (b) Model 2. Floating gate NVM structure with charge injection from top gate electrode. (c) Ferroelectric NVM structure. } \label{fig:models}
\end{figure}

In traditional NVM devices, the charges are injected from substrate/source, as shown in Fig.~\ref{fig:models}(a) (Model 1). Insulator 2 acts as a potential barrier in charge injection, for which lower barrier and thinner thickness will benefit the injection efficiency. However, in this structure, insulator 2 also acts as gate dielectric. For this function, the insulation of insulator 2 is crucial for the device performance, which demands higher barrier and thicker thickness. From above statement, we can find that there exists a conflicting requirement for insulator 2. This arises from the dual functions of insulator 2. This conflict becomes severe as the device size approaches the physical limit. \cite{houdt2011}

To overcome this difficulty, we propose a novel structure, in which the charges are injected from top gate electrode, as shown in Fig.~\ref{fig:models}(b) (Model 2). The insulator 1 acts as tunneling potential barrier and insulator 2 acts as gate dielectric. In this structure, each insulator layer plays one single function, thus they can be optimized respectively.

\section{Structure and formulae}

\begin{figure}[b]
\centering
\includegraphics[width=0.25\textwidth]{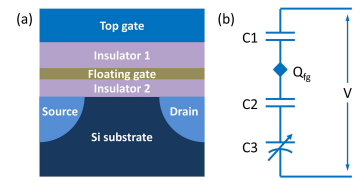}
\caption{(a) Layer structure of a typical NVM device. (b) Schematic diagram of the electronic circuit to model the device structure of (a).} \label{fig:circuit}
\end{figure}

A typical structure of floating gate NVM devices is shown in Fig.~\ref{fig:circuit}(a). The equivalent electronic circuit is depicted in Fig.~\ref{fig:circuit}(b). Each insulator layer is characterized by one capacitor. The capacitance between insulator 1 and the substrate is characterized by a variable capacitor depending on the interface state. The floating gate is represented by stored charge $Q_{fg}$. The band bending is neglected for simplicity when $V$ and $Q_{fg}$ are zeros.

For usable memory devices, the leakage is generally in low levels. For theoretical analysis, we assume the resistance of the structure is large enough to be ignored, while in real cases it should be critically addressed.

According to Kirchhoff's voltage law, we obtain Equ.~\ref{equ:cs1}.
\begin{equation}
V_1+V_2+V_3=V
\label{equ:cs1}
\end{equation}
where $V_1$, $V_2$ and $V_3$ are the voltage applied on $C_1$, $C_2$ and $C_3$, respectively. $V$ is the total voltage.
According to charge conservation, we obtain Equs.~(\ref{equ:cs2}) and (\ref{equ:cs3}).
\begin{equation}
V_2C_2=V_1C_1+Q_{fg}
\label{equ:cs2}
\end{equation}
\begin{equation}
V_2C_2=V_3C_3
\label{equ:cs3}
\end{equation}
By combining Equs.~(\ref{equ:cs1})-(\ref{equ:cs3}), the voltages applied on each layer and the interface between insulator 2 and substrate can be resolved, as shown in Equs.~(\ref{equ:cs4})-(\ref{equ:cs6}).
\begin{equation}
V_1=\frac{VC_2C_3-Q_{fg}(C_2+C_3)}{C_1C_2+C_2C_3+C_3C_1}
\label{equ:cs4}
\end{equation}
\begin{equation}
V_2=\frac{(VC_1+Q_{fg})C_3}{C_1C_2+C_2C_3+C_3C_1}
\label{equ:cs5}
\end{equation}
\begin{equation}
V_3=\frac{(VC_1+Q_{fg})C_2}{C_1C_2+C_2C_3+C_3C_1}
\label{equ:cs6}
\end{equation}
The following analysis is based on the Equs.~(\ref{equ:cs4})-(\ref{equ:cs6}), to discuss how to optimize memory window, electric field on the two insulator layers, and the influence of stored charges.

\section{Optimization conditions}
In this section, we discuss the requirements on optimizing the structure. Three aspects are discussed, which are essential for NVM performance. They are memory window, charge injection efficiency and the effect of stored charges.
\subsection{Memory window}
The memory window can be defined by the flat band voltage shift due to charge storage. For NVM structure, flat band means the band diagram of insulator 2 is flat, i.e. $V_2=0$. From Equ.~(\ref{equ:cs5}), by combining  $V_2=0$, we obtain $V=-\frac{Q_{fg}}{C_1}$. In general, we assume positive and negative charges can all be stored in floating gate. Then the memory window can be expressed as follows,
\begin{equation}
V_{window}=\frac{|Q_1|+|Q_2|}{C_1}
\label{equ:mw}
\end{equation}
where $Q_1$ and $Q_2$ are the stored charge in the floating gate by positive and negative charge injection, respectively. From Equ.~(\ref{equ:mw}), $V_{window}$ is proportional to $\frac{1}{C_1}$. Thus, the memory window can be tuned by the capacitance of insulator 1. Generally, for a large memory window, small $C_1$ is needed.

\subsection{Charge injection efficiency}
\label{sec:cje}
The efficiency of charge injection is dependent on the electric field applied on tunneling barrier (insulator 1 or insulator 2) in F-N tunneling mechanism. \cite{lezlinger1969} To elucidate the tunneling efficiency, from Equs.~(\ref{equ:cs4}) and (\ref{equ:cs5}), we deduced Equ.~(\ref{equ:ef}).
\begin{equation}
\frac{\partial{E_1}}{\partial{V}}/\frac{\partial{E_2}}{\partial{V}}=\frac{\partial{V_1}}{d_1\partial{V}}/\frac{\partial{V_2}}{d_2\partial{V}}=\frac{\epsilon_2}{\epsilon_1}
\label{equ:ef}
\end{equation}
where $E_1$, $E_2$, $d_1$, $d_2$, $\epsilon_1$ and $\epsilon_2$ are the electric fields, thicknesses and dielectric constants of insulator 1 and insulator 2, respectively.

From Equ.~(\ref{equ:ef}), the ratio of electric fields induced by external voltage in insulator 1 and insulator 2 is proportional to $\frac{\epsilon_2}{\epsilon_1}$. Assuming insulator 1 is tunneling barrier, then the higher dielectric constant ratio of insulator 2 to insulator 1, the higher the tunneling efficiency. Similar result can be deduced for insulator 2 as tunneling barrier. High electric field imposed on insulator 2 as in Model 1 may shorten the device lifetime due to oxide breakdown. \cite{lombardo2005}

\subsection{Stored charge}
From Equs.~(\ref{equ:cs4})-(\ref{equ:cs6}), for $Q_{fg}$ induced voltages, the voltage applied on insulator 1 equals to the summation of the voltage on insulator 2 and that on inverse layer, but with opposite signs. The voltage applied on the inversion layer determines the state of the transistor to be on or off. Then the higher the $Q_{fg}$ induced voltage on inverse layer  is, the higher the performance will be obtained. From Equs.~(\ref{equ:cs5}) and (\ref{equ:cs6}), Equ.~(\ref{equ:qfg}) is deduced.
\begin{equation}
\frac{\partial{V_3}}{\partial{Q_{fg}}}/\frac{\partial{V_2}}{\partial{Q_{fg}}}=\frac{C_2}{C_3}
\label{equ:qfg}
\end{equation}
When the device is in accumulation or flat band state, no inversion layer exists. $C_3\rightarrow\infty$. All the $Q_{fg}$ induced voltage is applied on insulator 2. It is trivial and the transistor is off. When the device is in inversion, $C_3$ is finite, which is determined by the semiconducting substrate largely. According to Equ.~(\ref{equ:qfg}), the ratio of $Q_{fg}$ induced voltages on inverse layer and insulator 2 equals to $\frac{C_2}{C_3}$. To obtain strong inversion, the larger value of Equ.~(\ref{equ:qfg}) is needed, which demands high value of $C_2$.
\subsection{Discussion}
\begin{table}[t]
\begin{tabular}{ccccc}
\hline
& \multicolumn{2}{c}{Model 1}&\multicolumn{2}{c}{Model 2}\\
parameter&insulator 1&insulator 2&insulator 1&insulator 2\\ \hline
$\epsilon$ &high & low& low & high \\
$d$&thick&thin&thin&thick\\
$C$&small&large&small&large\\
$E$&small&large&large&small\\ \hline
\end{tabular}
\caption{The requirements for the electric properties of materials and film thickness. $\epsilon$, $d$, $C$ and $E$ are dielectric constant, film thickness, capacitance and electric field, respectively.}
\label{tab:requirement}
\end{table}

According to each criteria above, the requirements for materials and film thicknesses can be summarized in Table~\ref{tab:requirement}. The contradiction discussed in Section~\ref{sec:contra} can be clearly seen. For Model 1, $\epsilon_2$ is low, $d_2$ is thin. To obtain large $C_2$, $d_2$ will be pushed to extreme thin. This contradiction compromises of the device performance. The situation becomes severe when the size of devices approaches extreme and cannot be conquered finally. For insulator 1, similar situation is faced. However, in Model 2, the contradiction vanishes. High $\epsilon$, large $C$ and thick $d$ for insulator 2 can be satisfied at the same time. And it is the same for insulator 1.

Ferroelectric NVM is deemed as a candidate to replace the current workhorse, floating gate NVM. However, according to our analysis, this would set high requirements for the ferroelectric materials. A typical ferroelectric NVM structure is shown in Fig.~\ref{fig:models}(c), \cite{setter2006} which is similar to Model 2. The electric dipole moment induced by external voltage is equivalent to charge injection from top gate electrode. Then the analysis on Model 2 can also be applied. Accordingly, the requirements for ferroelectric NVM are the same as those in Model 2. To efficiently change the dipole moment, the higher the electric field applied on the ferroelectric film is, the better the performance. However, as we know the dielectric constant of ferroelectric materials is often very high, e.g. several thousand for Pb(Zr$_x$Ti$_{1-x}$)O$_3$ and BaTiO$_3$ at room temperature. This lowers the efficiency of electric field applied on it according to Section~\ref{sec:cje}. If a high voltage is applied, the electric field in insulator 2 will be very high, which makes it easy to break down. \cite{lombardo2005} To avoid this, insulator 2 can be simply removed. However, the depolarizing electric field under direct contact of ferroelectric and semiconductor significantly decrease retention time. \cite{mehta1973}

The proposed structure is very similar to the traditional one, so that many other technologies, which are to improve NVM performance, can be applied without difficulties, such as multi-level cell \cite{eitan2000} and 3D stacking technology. \cite{xie2010b}

\section{Material selection and structure growth}
%In this section, we discuss the availability of high-k materials and the feasibility of device fabrication.

\subsection{Material selection}
\begin{table}[t]
\begin{tabular}{ccccc}
\hline
Material&$\epsilon$&bandgap&CB offset&VB offset\\
{}&{}&(eV)&(eV)&(eV)\\\hline
Si&&1.1& & \\
SiO$_2$&3.9&9&3.2 &4.7 \\
HfO$_2$, doped HfO$_2$&$\sim$~25&$\sim$~6.0&1.4 &3.5 \\
HfSiO$_4$&11&6.5&2.8 &3.6 \\
Al$_2$O$_3$&9&8.8&2.8 &4.9 \\
SrTiO$_3$&2000&3.2&0 &2.1 \\
La$_2$O$_3$&30&6&2.3 &2.6 \\
Ta$_2$O$_5$&22&4.4&0.35 &2.95 \\
TiO$_2$&80&3.5&0 &2.4 \\
Y$_2$O$_3$&15&6&2.3&2.6\\
ZrO$_2$&25&5.8&1.5 &3.2 \\
Si$_3$N$_4$&7&5.3&2.8 &1.4 \\
a-LaAlO$_3$&30&5.6&1.8 &2.7 \\
\hline
\end{tabular}
\caption{Properties of high-k materials. \cite{robertson2004} The CB and VB offsets are relative to Si. The data of Si and SiO$_2$ are listed for comparison.}
\label{tab:highk}
\end{table}

In practical devices, nanoparticle is ideal for charge storage medium, for its size suitable for small scale device fabrication. Narrow bandgap material particles, such as Si nanoparticles, are desired to use as floating gate material. SiO$_2$ can be used as low-k material for its excellent stability and insulation. There are many high-k materials can be selected, which have been extensively studied and are listed on Table~\ref{tab:highk}. \cite{robertson2004} From Table~\ref{tab:highk}, it is clear that dielectric constant, bandgap, conduction band (CB) offset and valance band (VB) offset distribute widely. Because high-k material acts as gate dielectric, charge tunneling is not desired. Thus large CB/VB band offset is desired. From Table~\ref{tab:highk}, except SrTiO$_3$, Ta$_2$O$_5$ and TiO$_2$, the CB/VB offsets are larger than 1.4 eV for all other materials. This makes it easy to choose the fitted material to design NVM devices.

\subsection{Feasibility of structure growth}
Nowadays, film growth techniques have been developed very well. Among them, MOCVD, MBE, PLD, etc. can be chosen for their reliable high film quality and reproducibility. The problems, which may influence the device performance, are the quality of narrow bandgap material nanoparticles inserted between high-k and low-k dielectric layers.

\section{Conclusions}
In summary, we pointed out the difficulty on scaling down the device size in NVM and proposed a new structure to overcome it. By analysis, it is possible to use this structure to achieve scaling down, low power consumption, high writing speed and long retention time.

\section{Postscript}
This idea was inspired by my previous work.\cite{cui2014apl,cui2014mrx}

%\bibliographystyle{IEEEtran}
%\bibliography{ref}

\end{document}